\documentclass[aps,twocolumn,english,showpacs]{revtex4}
\usepackage[T1]{fontenc}
\usepackage[latin9]{inputenc}
\usepackage{color}
\usepackage{amsmath}
\usepackage{graphicx}
\usepackage{amssymb}

\makeatletter
\@ifundefined{textcolor}{}
{%
 \definecolor{BLACK}{gray}{0}
 \definecolor{WHITE}{gray}{1}
 \definecolor{RED}{rgb}{1,0,0}
 \definecolor{GREEN}{rgb}{0,1,0}
 \definecolor{BLUE}{rgb}{0,0,1}
 \definecolor{CYAN}{cmyk}{1,0,0,0}
 \definecolor{MAGENTA}{cmyk}{0,1,0,0}
 \definecolor{YELLOW}{cmyk}{0,0,1,0}
 }

\makeatother

\usepackage{babel}

\begin{document}

\title{A Frustrated 3-Dimensional Antiferromagnet: Stacked $J_{1}-J_{2}$
Layers}

\author{Onofre Rojas\footnote{Departamento de Ciências Exatas, Universidade Federal de Lavras, CP 3037, 37200-000, Lavras, MG, Brazil.},
C. J. Hamer and J. Oitmaa}

\affiliation{School of Physics, The University of New South Wales, Sydney 2052,
Australia.}
\begin{abstract}
We study a frustrated 3D antiferromagnet of stacked $J_1 - J_2$ layers.
The intermediate 'quantum spin liquid' phase, present in the 2D case,
narrows with increasing interlayer coupling and vanishes at a triple
point. Beyond this there is a direct first-order transition from N{\'
e}el to columnar order. Possible applications to real materials are
discussed.
\end{abstract}
\pacs{PACS Indices: 05.30.-d,75.10.-b,75.10.Jm,75.30.Ds,75.30.Kz \\
\\  \\
(Submitted to  Phys. Rev. B) }
\maketitle

\section{Introduction}
\label{sec1}

The study of frustrated quantum antiferromagnets remains an active
field, characterized by strong interplay between theory and experiment.
A model which has been studied extensively (see \cite{chandra,dagotto,schulz,
richter,oweihong,bishop,singh,capriotti,siurak,sushkov,capriotti3,capriotti2,
singh2,roscilde,sirker}
and references therein)
is the so called `$J_{1}-J_{2}$
model', a spin-1/2 system on the two dimensional square lattice with
nearest and next-nearest neighbor interactions of strength $J_{1}$
and $J_{2}$ respectively, both being antiferromagnetic. The Néel
order at $T=0$, which pertains for $J_{2}=0$, is destabilized by
the frustrating $J_{2}$ interaction and vanishes at around $J_{2}/J_{1}\simeq
0.4$.
In the opposite limit, of large $J_{2}$, a columnar (often ambiguously
termed `collinear') ordered phase occurs, in which successive columns
(or rows) of spins alternate in direction. The columnar phase 
becomes unstable at $J_{2}/J_{1}\simeq0.6$. A magnetically
disordered region thus exists in the region $0.4\lesssim J_{2}/J_{1}\lesssim0.6$,
the nature of which remains not fully resolved.

It is of interest to ask what happens to this intermediate phase,
and indeed to the entire phase diagram, when these $J_{1}-J_{2}$
layers are coupled, perhaps weakly, in the third direction, forming
a 3-dimensional structure. 
This is the main topic of this paper. 
It has already been studied by Schmalfuss {\it et al.} \cite{schmalfuss} using
the coupled-cluster and rotation-invariant Green's function methods. 
They found that the disordered region becomes narrower as a function of $J_2$
when $J_3$ is increased, and vanishes entirely at $J_3/J_1 \simeq 0.2 - 0.3$.
We will
address the question using series expansion methods \cite{ohweihong} in both Néel
and columnar phases, as well as first order spin-wave theory.

It has been argued recently that the layered materials $\mathrm{Li_{2}VOSiO_{4}}$,
$\mathrm{Li_{2}VOGeO_{4}}$ are well represented by the spin-1/2 $J_{1}-J_{2}$
model with $J_{2}\gg J_{1}$, i.e. in the columnar phase. In ref.
\cite{Rosner} high-temperature expansions for the specific heat and
magnetic susceptibility for the purely 2-d model were used to try
to constrain the values of the exchange parameters by fitting to the
experimental data. At the same time an {\it ab initio} local density approximation
(LDA) calculation of the exchange parameters $J_{1}$, $J_{2}$, $J_{3}$
yielded (0.75K, 8.8K, 0.25K) for the Si material and (1.7K, 8.1K,
0.19K) for the Ge system. This
 suggests that the coupling in the third direction is by no means
negligible and ought to be included in a fitting procedure. 

We mention also that a popular, though not universally accepted, scenario
to understand the magnetic properties of the recently discovered superconducting
iron pnictides\textcolor{red}{{} }\textcolor{black}{is via a spin-1
layered
$J_{1}-J_{2}$ model \cite{Uhrig,Holt}. While we do not consider these systems
explicitly here, our results may have some relevance.}

\begin{figure*}
\includegraphics[scale=0.85]{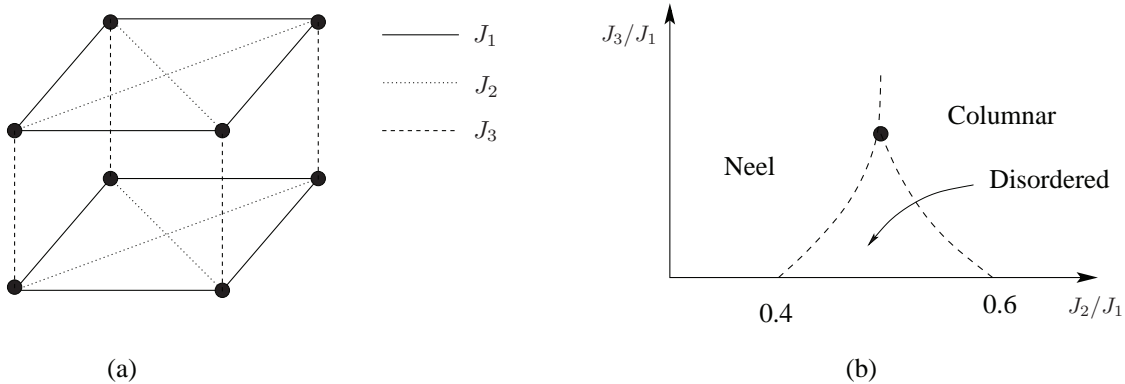}
\caption{(a) Tetragonal unit cell of the model; (b) Conjectured
phase diagram of the model at $T=0$. The transition lines are (probably)
first-order and the solid circle is thus a triple point. }
\label{fig1}
\end{figure*}

Our model is a 3-dimensional spin-1/2 antiferromagnet, on a tetragonal
lattice, as shown in Figure \ref{fig1}(a). Frustrating $J_{2}$ interactions
occur in the x-y plane but not in the third direction. All interactions
are antiferromagnetic.

The Hamiltonian, in standard notation, is

\begin{equation}
H=J_{1}\sum_{<ij>}^{(1)}\mathbf{S}_{i}.\mathbf{S}_{j}+J_{2}\sum_{<ik>}^{(2)}\mathbf{S}_{i}.\mathbf{S}_{k}+J_{1}\sum_{<il>}^{(3)}\mathbf{S}_{i}.\mathbf{S}_{l}\end{equation}
where the summations are over the three classes of coupling respectively.

In the following section, section \ref{sec2}, we present our zero temperature
calculations and results, for the ground state energy and sublattice
magnetization (order parameter), for both the Néel and columnar phases.
Our results support a phase diagram of the form shown in Figure \ref{fig1}(b),
and we estimate the location of the triple point at $J_{2}/J_{1}=0.54(3)$
and $J_{3}/J_{1}=0.16(3)$ Our series results are compared with the
results of
linear spin-wave theory. In section \ref{sec3} we use series methods to compute
the 1-magnon dispersion curves, in both ordered phases, and again
compare our results with the spin-wave predictions. The spin-wave
theory follows standard lines and, for completeness, is outlined in the appendix.
In section \ref{sec4} we present our conclusions
and discussion.

\section{Ground State Bulk Properties}
\label{sec2}

The series expansion method is based on perturbative calculations
for a sequence of finite connected clusters, which are then combined
to obtain a series for the bulk system. In practice it is possible
to treat of order $10^{6}$ different clusters and to obtain series
of order 10-20 terms, depending on the model. The Hamiltonian is written
in the standard form $H=H_{0}+\lambda V$ where $H_{0}$ has a simple
known ground state. In the present work we use an Ising expansion
in which $H_{0}$ consists of the diagonal $S_{i}^{z}S_{j}^{z}$ terms
and the quantum fluctuations are included in the perturbation $V$.
Series are obtained in powers of $\lambda$, and extrapolated to $\lambda=1$
via standard Padé or differential approximant methods. The interested
reader is referred to the book \cite{ohweihong} for more detail.

\begin{figure*}
\includegraphics[scale=0.4]{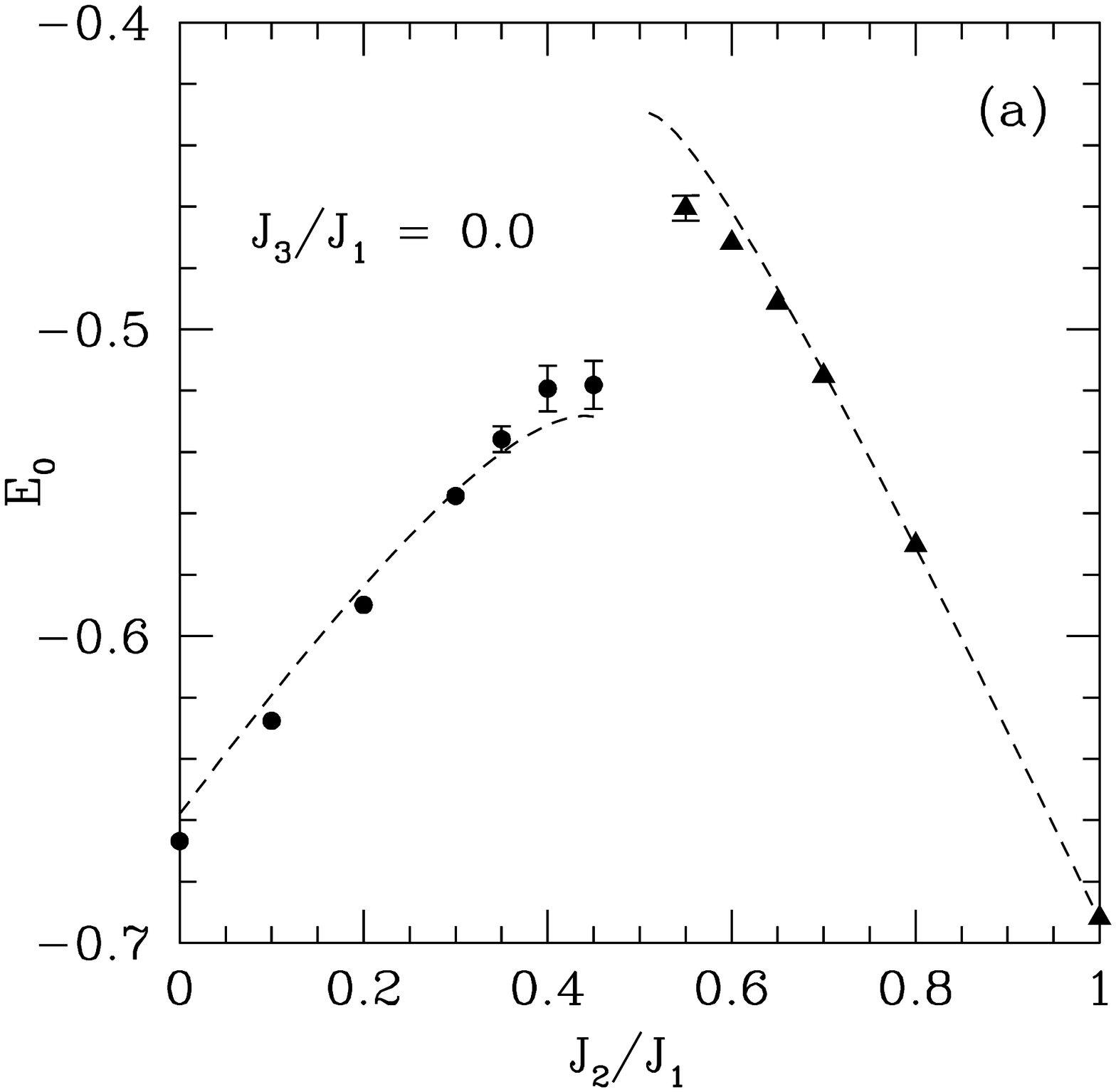}\includegraphics[scale=0.4]{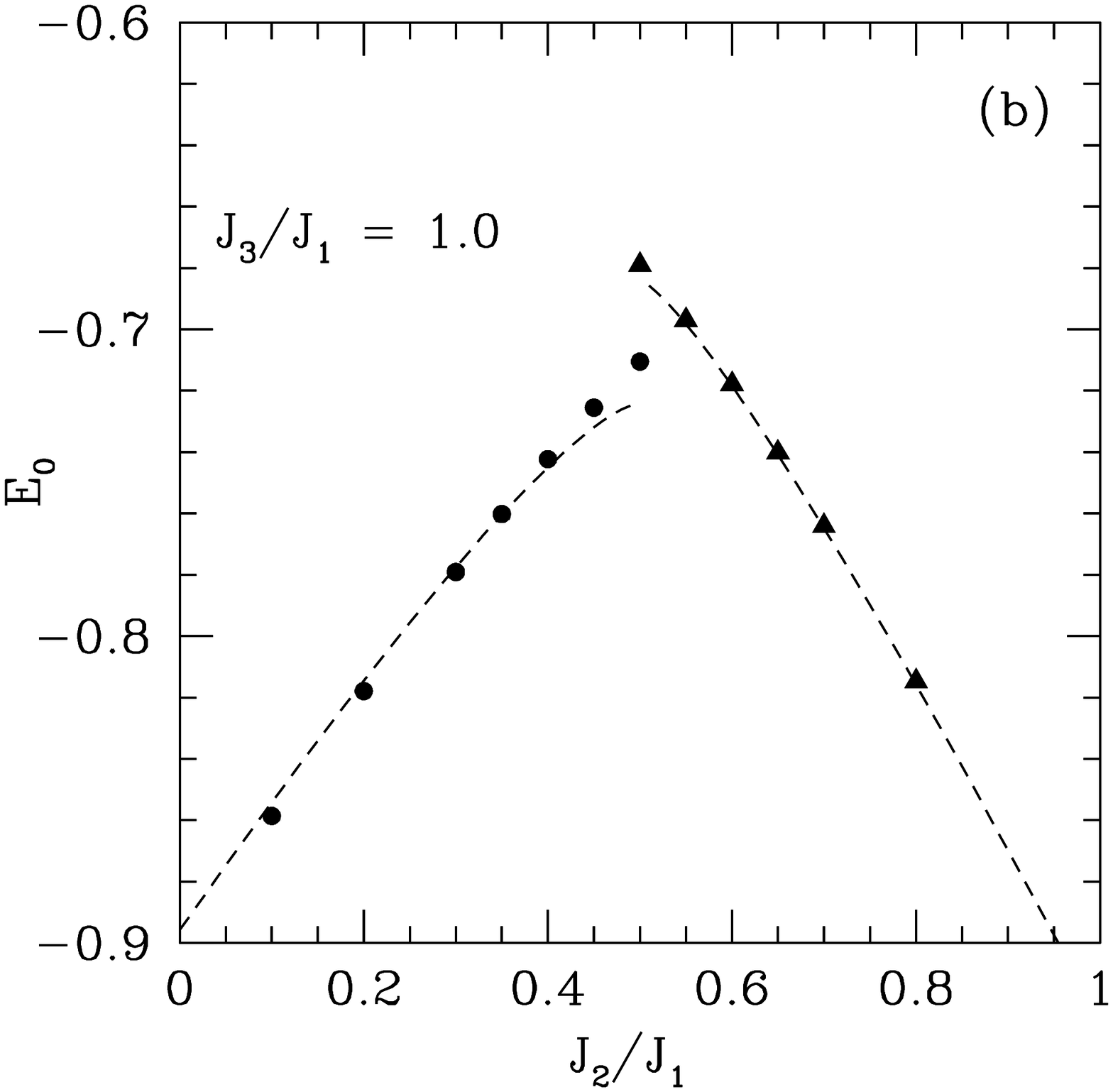}
\caption{Series estimates of the ground-state energy per site at couplings (a) $J_3/J_1 = 0.0$, (b) $J_3/J_1 = 1.0$. Filled circles
- N\'eel phase; filled triangles - columnar phase. The dashed lines are linear
spin-wave predictions in each phase.}
\label{fig2}
\end{figure*}
\begin{figure*}
\includegraphics[scale=0.4]{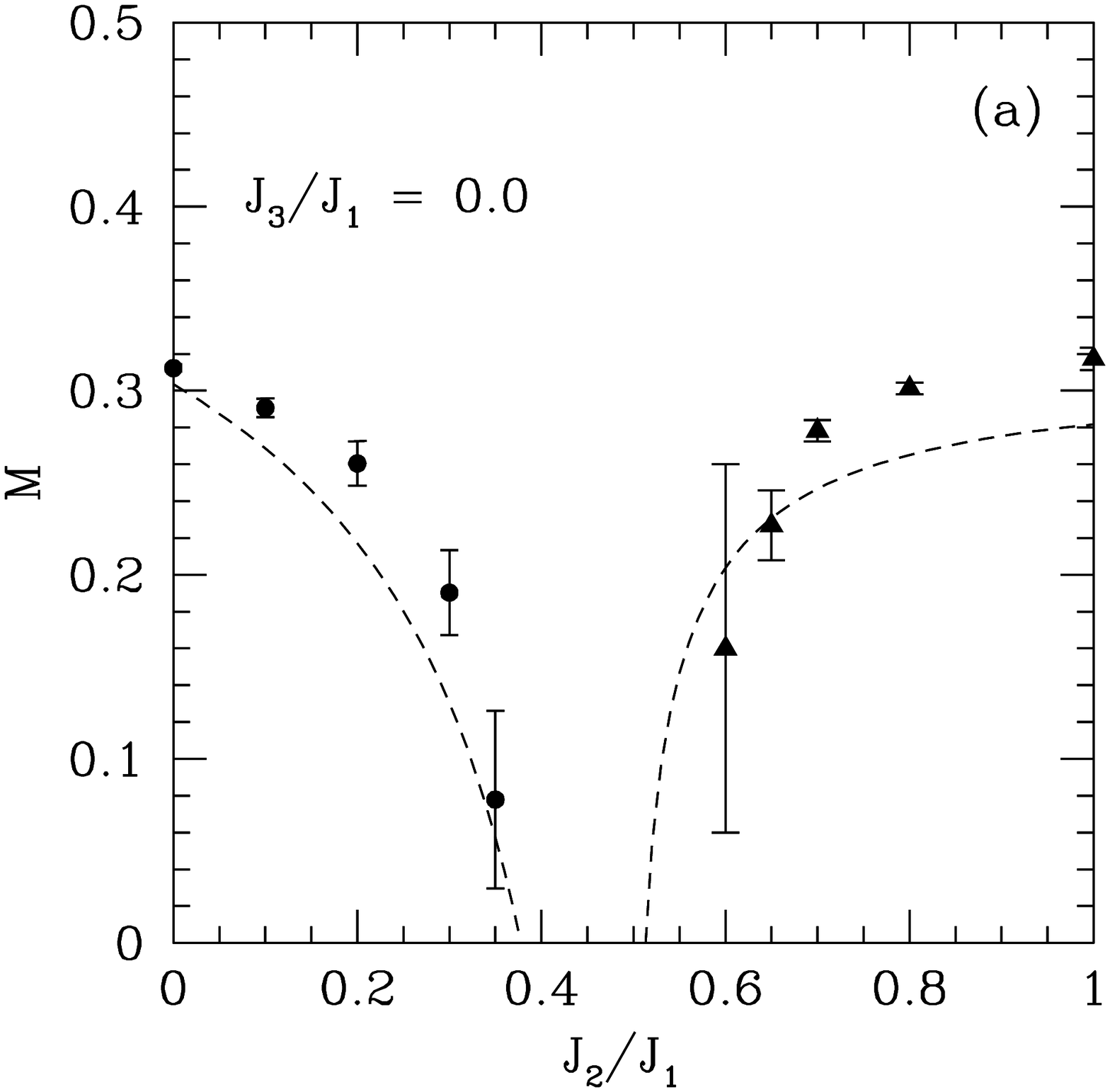}\includegraphics[scale=0.4]{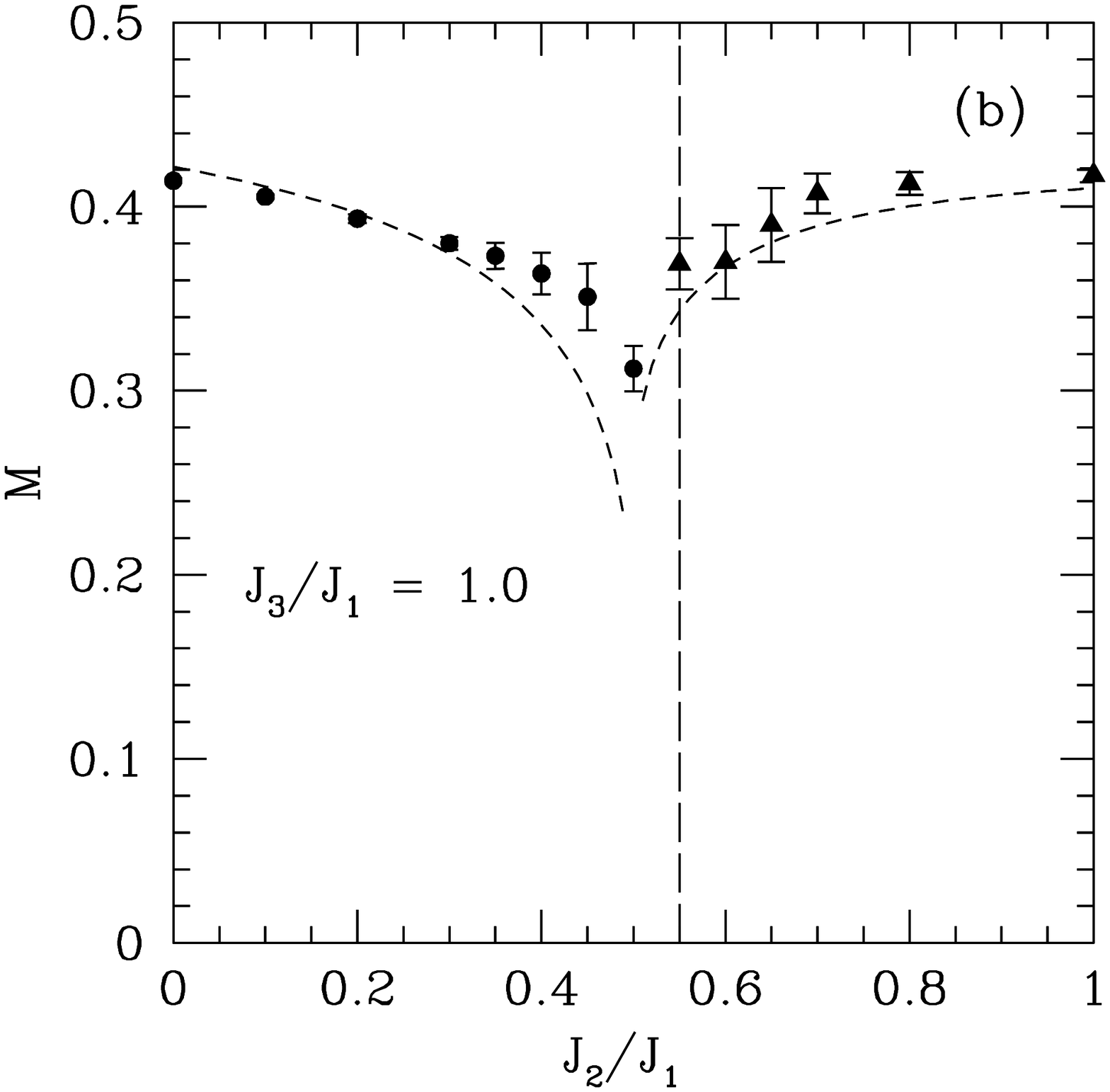}

\caption{
Series estimates of the staggered magnetizations in each phase at couplings (a)
$J_3/J_1 = 0.0$, (b) $J_3/J_1 = 1.0$. Filled circles - N\'eel phase; filled
triangles - columnar phase. The dashed curves are linear spin-wave
predictions in each phase. The dashed vertical line in Figure \ref{fig3}b marks
the estimated position of the first-order transition. 
}
\label{fig3}
\end{figure*}

\subsection{Néel Phase}

It is well known that the unfrustrated square lattice ($J_{2}=J_{3}=0$)
has reduced Néel order in the ground state. 
This is a 2-sublattice structure with spins pointing in opposite directions
on the two sublattices. For technical reasons it is useful to perform
a spin rotation on the B sublattice, making the unperturbed ground
state ferromagnetic. The Hamiltonian, for a lattice of $N$ sites
is then written as

\begin{equation}
H=-\frac{1}{4}\left(2J_{1}-2J_{2}+J_{3}\right)N+H_{0}+\lambda V\end{equation}
with 

\begin{widetext}
\begin{equation}
H_{0}  =  J_{1}\sum_{<ij>}^{(1)}\left(\frac{1}{4}-S_{i}^{z}S_{j}^{z}\right)+J_{2}\sum_{<ik>}^{(2)}\left(S_{i}^{z}S_{k}^{z}-\frac{1}{4}\right)+J_{3}\sum_{<il>}^{(3)}\left(\frac{1}{4}-S_{i}^{z}S_{l}^{z}\right)
\end{equation}

\begin{equation}
V=\frac{1}{2}J_{1}\sum_{<ij>}^{(1)}\left(S_{i}^{+}S_{j}^{+}+S_{i}^{-}S_{j}^{-}\right)+\frac{1}{2}J_{2}\sum_{<ik>}^{(2)}\left(S_{i}^{+}S_{k}^{-}+S_{i}^{-}S_{k}^{+}\right)+\frac{1}{2}J_{3}\sum_{<il>}^{(3)}\left(S_{i}^{+}S_{l}^{+}+S_{i}^{-}S_{l}^{-}\right)\end{equation}
\end{widetext}

We have added and subtracted constant terms to make the unperturbed
energy zero.

We have obtained series for the ground state energy $E_{0}$ and the
magnetization $M$ to order $\lambda^{9}$. The rapid proliferation
of clusters with 3 bond types (there are 320274 with 9 or fewer sites)
limits the length of the series obtainable. The data are far too
extensive to present, but can be provided on request. The values of
$E_0$ and $M$ can then be estimated, with some uncertainty, for any values
of $J_{2}$ and $J_{3}$. It is convenient to set $J_{1}=1$ to set
the energy scale. A display of the results is deferred to the third
subsection, where they are presented together with the results for
the columnar phase.

\subsection{Columnar Phase}

In the columnar phase the spins on alternating columns in a particular x-y plane
will point in opposite directions. This satisfies all of the strong
$J_{2}$ bonds but leaves half of the $J_{1}$ bonds frustrated. In
adjacent planes these are shifted by one lattice spacing, leaving all
of the $J_{3}$ bonds satisfied. This structure has an additional
2-fold degeneracy as columns can equally well be `rows'. Again it
is convenient to carry out a spin rotation on `down' sites, yielding

\begin{equation}
H=-\frac{1}{4}\left(2J_{2}+J_{3}\right)N+H_{0}+\lambda V\end{equation}

with 
 
\begin{widetext}
\begin{equation}
H_{0}=J_{1}\sum_{<ij>}^{(1x)}\left(\frac{1}{4}-S_{i}^{z}S_{j}^{z}\right)+J_{1}\sum_{<ij>}^{(1y)}\left(S_{i}^{z}S_{j}^{z}-\frac{1}{4}\right)+J_{2}\sum_{<ik>}^{(2)}\left(\frac{1}{4}-S_{i}^{z}S_{k}^{z}\right)+J_{3}\sum_{<il>}^{(3)}\left(\frac{1}{4}-S_{i}^{z}S_{l}^{z}\right)\end{equation}

\begin{align}
V= & \frac{1}{2}J_{1}\sum_{<ij>}^{(1x)}\left(S_{i}^{+}S_{j}^{+}+S_{i}^{-}S_{j}^{-}\right)+\frac{1}{2}J_{1}\sum_{<ij>}^{(1y)}\left(S_{i}^{+}S_{j}^{-}+S_{i}^{-}S_{j}^{+}\right)\nonumber \\
 & +\frac{1}{2}J_{2}\sum_{<ik>}^{(2)}\left(S_{i}^{+}S_{k}^{-}+S_{i}^{-}S_{k}^{+}\right)+\frac{1}{2}J_{3}\sum_{<il>}^{(3)}\left(S_{i}^{+}S_{l}^{+}+S_{i}^{-}S_{l}^{-}\right)\end{align}
\end{widetext}
Here the notations $(1x)$ and $(1y)$ refer to nearest neighbor $J_{1}$
bonds in the $x$ and $y$ direction in a plane, respectively.
We have calculated the expansion to order $\lambda^{8}$ inclusive,
involving a total of 114650 distinct clusters with 4 types of bond.

\begin{figure}
\includegraphics[scale=0.4]{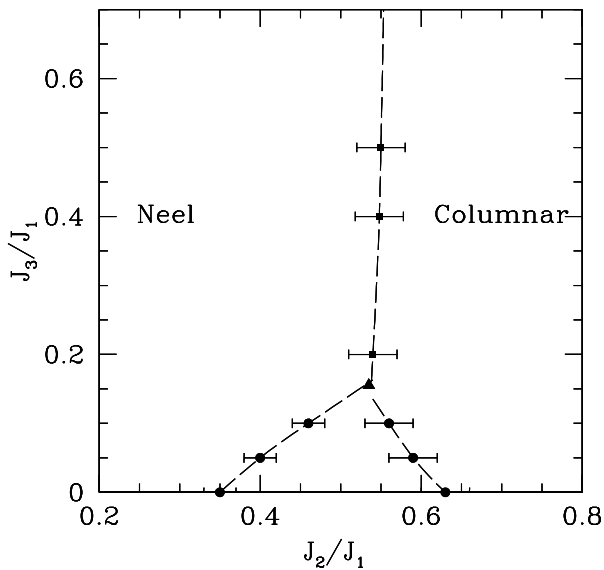}

\caption{
Phase diagram obtained from various series expansions. The dashed lines
are first-order transitions, and the filled triangle is a triple point.
}
\label{fig4}
\end{figure}

\begin{figure*}
\includegraphics[trim = 0mm 0mm 0mm 0mm,scale=0.75,clip]{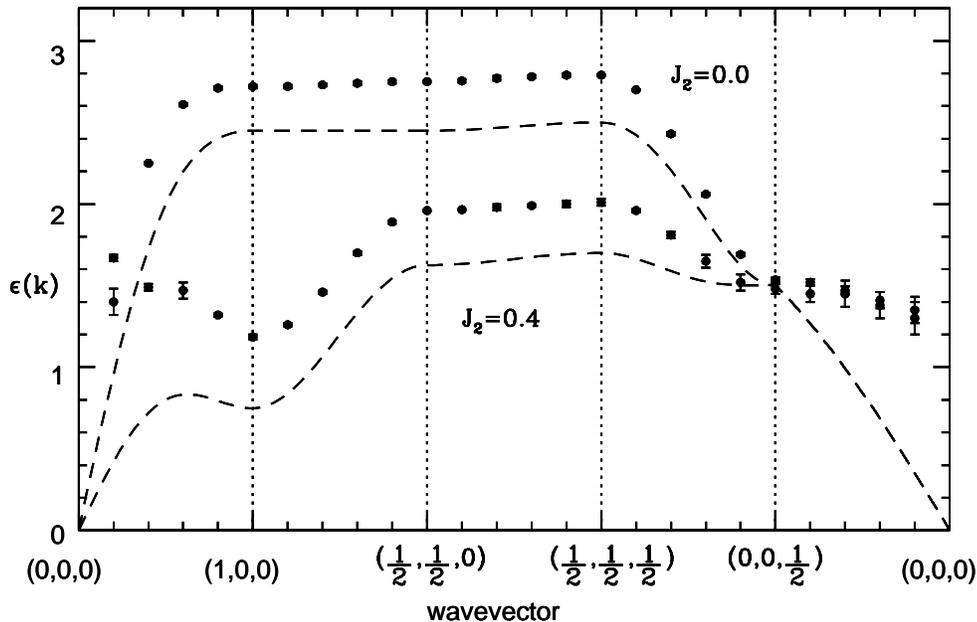}
\caption{
Series estimates of the single-magnon dispersion relations in the N{\' e}el
phase at  $J_2/J_1 = 0.0, J_3/J_1 = 0.5$ (upper curve), and
  $J_2/J_1 = 0.4, J_3/J_1 = 0.5$ (lower curve). The dashed lines are linear spin-wave predictions.
Values of ${\bf q} = {\bf k}/\pi$ along the path are shown at the bottom.
}
\label{fig5}
\end{figure*}
\subsection{Results}

Series were computed and analysed for the cases $J_3 =
0,0.05,0.1,0.2,0.5,1.0$ and a range of values of $J_2$ in both the
N\'eel and columnar phases. We set $J_1 = 1$ throughout.

Figure 2 shows estimates of the ground state energy per site versus $J_2$
for two cases, $J_3 = 0$ (Fig.2a) and $J_3 = 1.0$ (Fig.2b).The energy is
evaluated by forming Pad\'e approximants to the series and evaluating
these at $\lambda = 1$. Error bars, where shown, represent confidence
limits, based on the degree of consistency between high-order
approximants.

As is apparent, the behaviour of the energy estimates is quite different
in the two cases. For $J_3 = 0$, i.e the $J_1 - J_2$ square lattice
model, the energy estimates in the intermediate range $J_2 = (0.4,0.6)$
become erratic and the two curves, from the N\'eel and columnar series
respectively, do not appear to join. However, for $J_3 = 1$, the
estimates are quite precise and the two curves cross near $J_2 \simeq
0.55$. The clear difference in slope of the two branches at the crossing
point indicates a direct first-order transition between the two phases.

To analyse the magnetization series we first performed a Huse
transformation \cite{huse1988} to a new variable $x = 1
-\sqrt{1-\lambda}$ 
to remove the square-root singularity at $\lambda = 1$, expected from
spin-wave theory. This procedure has been used in earlier work on the 2D
antiferromagnet \cite{huse1988,zheng1991}. Pad\'e approximants to the
new series were then evaluated at $x = 1$.

The results are shown in Figure 3 and again display a striking difference
between the two cases $J_3 =0$ and $J_3 = 1$. For $J_3 = 0$ we see clear
evidence of an intermediate non-magnetic phase, with N\'eel and columnar
magnetizations vanishing near $J_2 \simeq 0.4$ and $J_2 \simeq 0.6$
respectively. The error bars are large near the transition points and it
is not possible to say whether the transitions are first- or
second-order. It is generally believed that the columnar-disordered
transition is first-order while the N\'eel-disordered transition appears
second-order but may, in fact, be weakly first-order \cite{sirker}.
On the other hand, for $J_3 = 1$ (Fig.3b) it is clear that both
magnetizations remain finite throughout their phase. This is consistent
with a direct first-order transition.

Proceeding in this way with other values of $J_3$, we construct a phase
diagram as shown in Figure 4, which confirms the schematic phase diagram
in Figure 1b. As is apparent, the disordered phase narrows as the
interplane coupling $J_3$ is increased, and vanishes at the estimated
point $J_2/J_1 = 0.54 \pm 0.03, J_3/J_1 = 0.16 \pm 0.03$. This value of
$J_3$ is a little lower than found previously \cite{schmalfuss}.

\begin{figure*}
\includegraphics[scale=0.8]{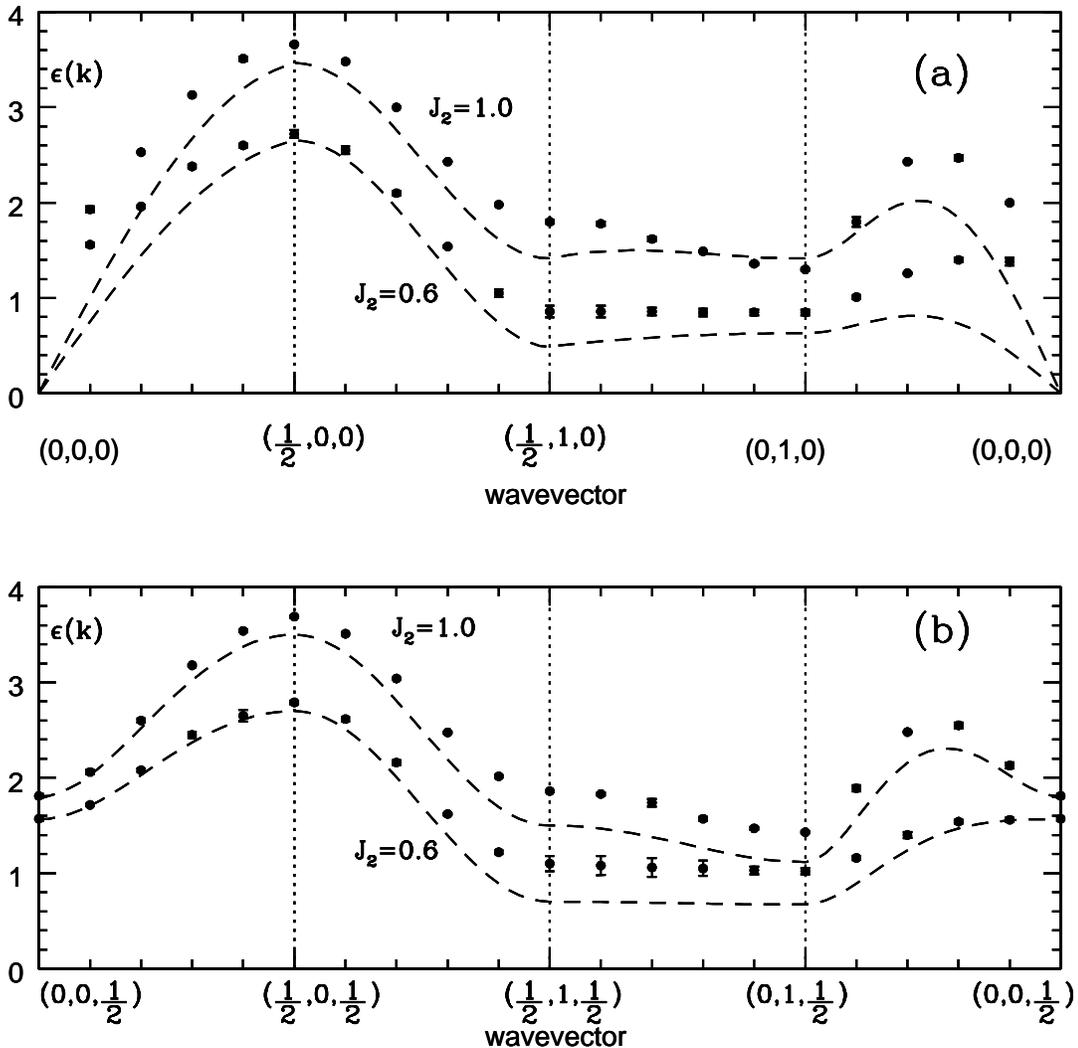}
\caption{
Series estimates of the single-magnon dispersion relations in the columnar
phase at (a) $J_2/J_1 = 1.0, J_3/J_1 = 0.5, k_z = 0 $ (upper curve) and
 $J_2/J_1 = 0.6, J_3/J_1 = 0.5, k_z =0 $ (lower curve), and (b) $J_2/J_1 = 1.0, 
 J_3/J_1 = 0.5, k_z = \pi/2$, (upper curve) and $J_2/J_1 = 0.6, J_3/J_1
= 0.5, k_z = \pi/2$ (lower curve). The dashes lines are linear spin-wave
predictions.
Values of ${\bf q} = {\bf k}/\pi$ along the path are shown at the bottom.
}
\label{fig6}
\end{figure*}

\section{Excitations}
\label{sec3}

It is also of interest to investigate the spectrum of single-magnon
excitations in the ordered phases and to see how these change on
increasing the interlayer coupling $J_3$. Standard series methods exist
for calculating the excitation energy throughout the Brillouin zone
\cite{ohweihong}, and we utilize these in the following.

\begin{figure*}
\includegraphics[scale=0.4]{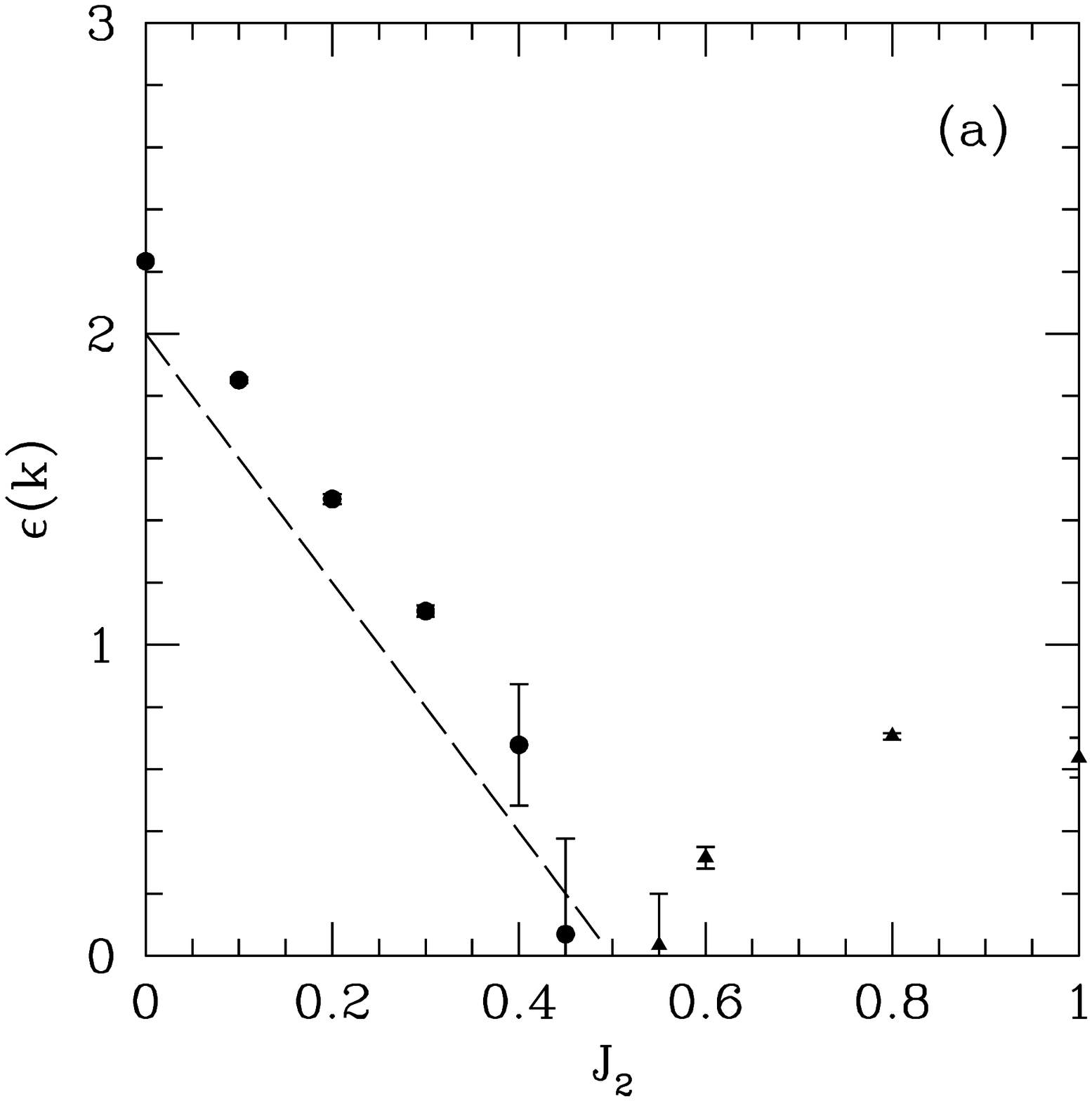}\includegraphics[scale=0.4]{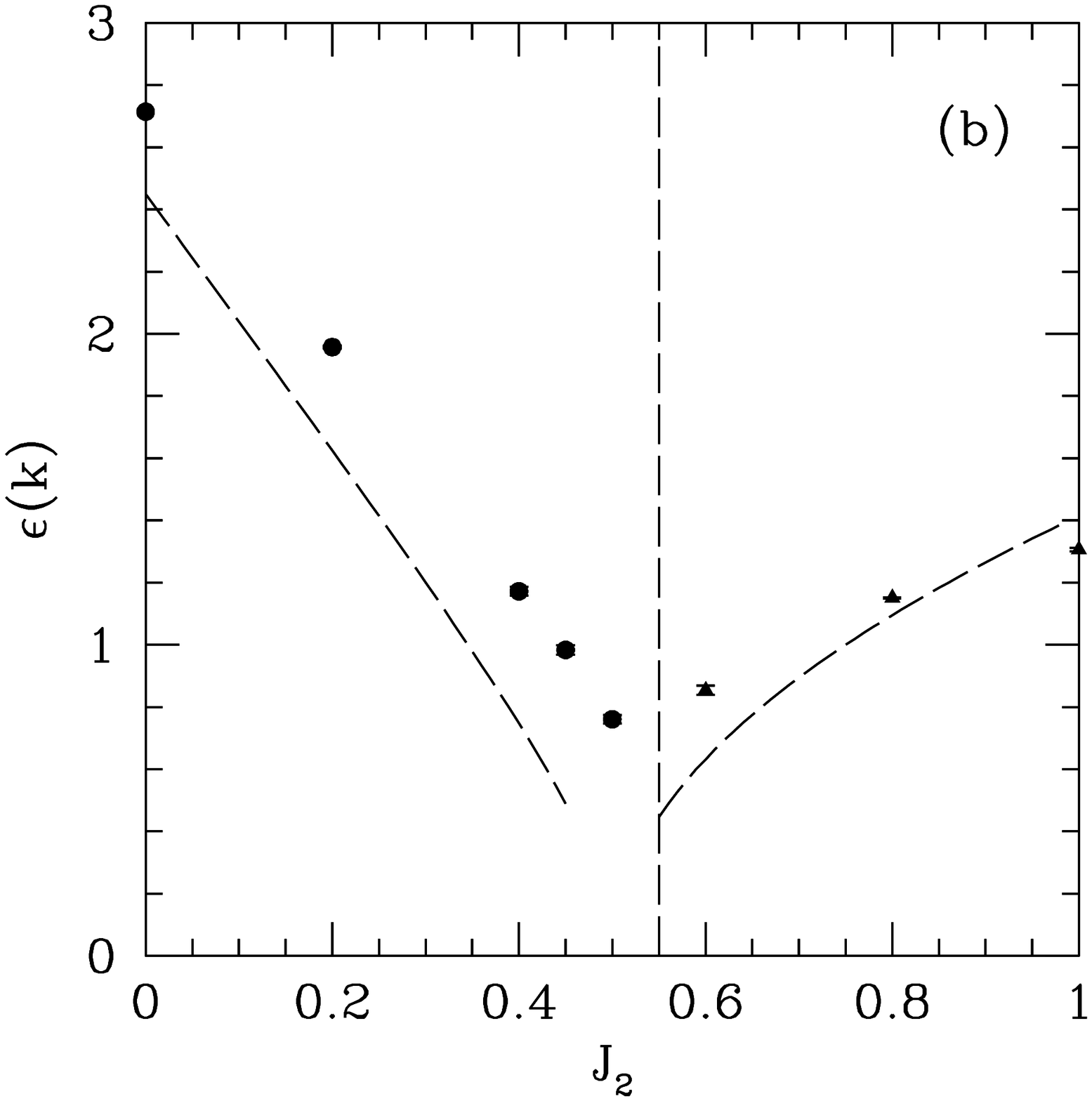}
\caption{
Series estimates of the energy gap $\epsilon(\bf{k})$ at ${\bf k} =
 (0,\pi,0)$ as a function of $J_2$ at (a) $J_3 = 0$
and (b) $J_3 = 0.5$. Filled circles - N{\' e}el phase; filled triangles -
columnar phase. The dashed lines are linear spin-wave
predictions in each phase.}
\label{fig7}
\end{figure*}

\subsection{N{\' e}el Phase}

Series were obtained to order $\lambda^7$, based on 255196 distinct
clusters of up to 8 sites. In Fig. 5 we show dispersion curves
for two values of the frustrating interaction $J_2 = 0.0, 0.4$ for $J_3
= 0.5$, along various symmetry lines in the Brillouin zone, as shown. 
For comparison we also show the results of linear spin-wave theory.

Several features are worth noting:
\begin{itemize}
\item[(1)] A very pronounced dip develops at wavevector $(\pi,0,0)$
(and, by symmetry, also at $(0,\pi,0)$) with increasing $J_2$. Linear
spin-wave theory has the gap vanishing at $J_2 = 0.5$ (see Appendix).
\item[(2)] There is a distinct `shoulder' at $(0,0,\pi/2)$, which is not
seen in linear spin-wave theory. Indeed, along the whole line
$(0,0,k_z)$ the energy depends only weakly on $J_2$.
\item[(3)] Linear spin-wave theory reproduces the overall dispersion
curves reasonably well, but systematically underestimates the excitation
energies.
\end{itemize}

We note also that the series become very irregular near the Goldstone
point $\bf{k} = 0$, and the Pad\'e approximants do not accurately show
the vanishing of the gap at this point. This has been noted in previous
work \cite{singh1993}, and an indirect way found to overcome this problem, which
also allows calculation of the spin-wave velocity. However, we do not
address this further in the present work.

\subsection{Columnar Phase}

Series were obtained to order $\lambda^7$, based on 490487 distinct
ckusters with up to 8 sites, and 4 possible bond types. In Figure
\ref{fig6} we show dispersion curves for $J_3 = 0.5$ and $J_2 = 1.0, 0.6$
along various symmetry lines in the Brillouin zone. The predictions of linear
spin-wave theory are shown as dashed lines. 

The most notable feature is the depressed excitation energy along the
line $(\pi/2,\pi,0)$ to $(0,\pi,0)$. The dispersion curve in this region
flattens as $J_2$ decreases from 1.0 to 0.6, in both cases (a) and (b).
According to linear spin-wave theory this whole branch becomes soft at
$J_2/J_1 = 0.5$, the classical transition point. Again linear spin-wave
theory tends to underestimate the excitation energy, but not as badly as
for the N\'eel phase. The same difficulty with the Goldstone point $
\bf{k} = 0$, discussed above for the N\'eel phase, occurs here.

\subsection{The Energy Gap at $(0,\pi,0)$}

As the phase transition is approached, from either the N\'eel or
columnar side, we expect increasing quantum fluctuations at wavevector $
\bf{k} = (0,\pi,0)$, which should be manifest through a decreasing
energy gap at this point. As mentioned above, linear spin-wave theory
predicts a vanishing excitation energy at this $\bf{k}$ point at the
classical transition point $J_2/J_1 = 0.5$.
Hence the variation of the gap is of some interest,
and our results are shown in Figure 7, as a function of $J_2$, for two
values of the coupling ratio $J_3/J_1 = 0,\ 0.5$.

 In the case
$J_3/J_1 = 0.5$ (Fig. \ref{fig7}b) the gaps remain of substantial size right up to the
direct first-order transition, and the gaps appear to be of roughly
equal magnitude in both phases at the transition. This is a further
sign that the transition between the two ordered phases is first
order at this value of $J_3$.

Figure \ref{fig7}a shows the energy gap at $J_3 = 0$. Note that
whereas linear spin wave theory has the gap vanishing at all
couplings in the columnar phase, it is in fact finite at larger
$J_2$, as given by a modified spin-wave theory and earlier series
results \cite{singh2}. Note also that
the energy gaps in both phases extrapolate to
zero before reaching the classical transition point $J_2/J_1 = 0.5$, providing
further evidence of a disordered intermediate phase when $J_3 = 0$.
Taken at face value, the results indicate that the gap remains
finite at the estimated transition points $J_2 \simeq 0.4$ and $J_2
\simeq 0.6$, which would indicate first order transitions. Naive
extrapolations are not entirely reliable near a critical point,
however, and a more precise check of this point would be of
interest. A Dlog Pad{\' e} analysis of the series in $\lambda$
gives no consistent estimates of the critical point in this case.

\section{conclusions}
\label{sec4}

We have used series expansion methods to study a system of stacked $J_1-J_2$
layers with antiferromagnetic coupling $J_3$ between the layers. This
model is applicable to the layered materials Li$_2$VOSiO$_4$ and
Li$_2$VOGeO$_4$, as well as being of intrinsic theoretical interest.
In agreement with earlier work of Schmalfuss {\it et el.} \cite{schmalfuss}, we
find that the disordered region of the phase diagram becomes narrower as $J_3$
is increased, and vanishes completely at a triple point, beyond which
there is a direct first-order transition between the N\'eel and columnar
phases. We estimate the location of the triple-point as $J_2/J_1 = 0.54
\pm 0.03, J_3/J_1 = 0.16 \pm 0.03$, the $J_3$ value being somewhat lower
than in ref. \cite{schmalfuss}. 

As well as mapping out the phase ground state phase diagram, we have
computed magnon dispersion curves along symmetry lines in the Brillouin
zone, in both the N\'eel and columnar phases, for various parameter
values. This would allow a more critical evaluation of the validity of
the model for the materials mentioned above, when experimental
inelastic neutron scattering results for magnon energies become
available.

After this work had been completed we became aware of two other recent
studies of this model. Nunes {\it et al.} \cite{nunes}, using an
effective field theory approach, concluded that the triple point (they
refer to this as a critical end-point, assuming the N{\' e}el to
disordered transition is second order) lies at $J_2/J_1 = 0.56, \
J_3/J_1 = 0.67$, values much higher than obtained in the present work
and in other work \cite{schmalfuss,Holt}. Majumdar \cite{majumdarnew} has carried out a
spin-wave calculation to second order ($1/S^2$) and concludes that the
intermediate phase remains present even at $J_3 = 1$, which seems
surprising in view of the other work referred to above.

Finally we mention another recent study \cite{majumdar} of a different
generalization of the 2-dimensional $J_1 - J_2$ model, in which the
frustrating $J_2$ interactions are included in all spatial directions.
That model appears also to have an interesting phase diagram, and could
be studied by series methods.

\section{Acknowledgements}
We are grateful for the computing resources provided by the Australian
Partnership for Advanced Computing (APAC) National Facility.

\appendix
\section{Linear Spin-Wave Theory}
\label{appendix}

In the Appendix we present
the results of linear spin-wave theory (LSWT) for the model, for both the
N{\' e}el and columnar phases. The results have been used in the main
text to compare with the more accurate series results.

The basic procedures of LSWT are well known so we will not include all
details.

\subsection{N{\' e}el Phase}

There are two equivalent sublattices A, B and two sets of bosons
describing spin deviations from the classical N{\' e}el state. After
Fourier transformation the Hamiltonian is, to quadratic order

\begin{widetext}
\begin{eqnarray}
H & = & -NS^2(2J_1-2J_2+J_3) +2S(2J_1-2J_2+J_3)\sum_k(a^{\dagger}_ka_k +
b^{\dagger}_kb_k) \nonumber \\
 & & +4SJ_1\sum_k\gamma_k(a^{\dagger}_kb^{\dagger}_k +
a_kb_k) +4SJ_2\sum_k \mu_k(a^{\dagger}_ka_k +
b^{\dagger}_kb_k) +2SJ_3\sum_k \nu_k(a^{\dagger}_kb^{\dagger}_k +    
a_kb_k)
\end{eqnarray}
\end{widetext}

with
\begin{equation}
\gamma_k = \frac{1}{2}(\cos k_x + \cos k_y),  \mu_k = \cos k_x \cos
k_y,  \nu_k = \cos k_z
\end{equation}
 
This is then diagonalized by a standard Bogoliubov transformation,
yielding

\begin{equation}
H = E_0 + \sum_k \omega_k (A^{\dagger}_kA_k +
B^{\dagger}_kB_k) 
\end{equation}
with
\begin{eqnarray}
\omega_k & = & 2S\sqrt{P^2_k - Q^2_k} \nonumber \\
P_k & = & 2J_1 + J_3 - 2J_2(1-\mu_k) \nonumber \\
Q_k & = & 2J_1\gamma_k +J_3\nu_k
\end{eqnarray}

The ground state energy is
\begin{equation}
E_0 = -NS(S+1)(2J_1-2J_2+J_3)+\sum_k \omega_k
\end{equation}
and the sublattice magnetization is
\begin{equation}
M = S+\frac{1}{2} - \frac{2S}{N}\sum_k \frac{P_k}{\omega_k}
\label{eq14}
\end{equation}
We note that for small $k$
\begin{eqnarray}
P_k + Q_k & = & 2(2J_1+ J_3)-\frac{1}{2}(J_1+2J_2)(k^2_x + k^2_y) \nonumber
\\
 & & -
\frac{1}{2}J_3 k^2_z + \cdots \nonumber \\
P_k - Q_k & = & (\frac{1}{2}J_1-J_2)(k^2_x + k^2_y) +     
\frac{1}{2}J_3 k^2_z + \cdots 
\end{eqnarray}
so there is a linear Goldstone mode at ${\bf k} = 0$, provided $J_2 < 
J_1/2$. The phase becomes unstable for $J_2 \ge J_1/2$.

Another special case occurs at ${\bf k} = (\pi,0,0)$. Let ${\bf k'} =
{\bf k} - (\pi,0,0)$ be small, then
\begin{eqnarray}
P_k + Q_k & = & 
2J_1-4J_2+\frac{J_3}{2}{k'}_z^2 +J_2({k'}_x^2+{k'}_y^2) + \cdots \nonumber \\
P_k - Q_k & = & 
2J_1-4J_2+2J_3+J_2({k'}_x^2+{k'}_y^2) \nonumber \\
 & & -\frac{J_3}{2}{k'}_z^2 + \cdots 
\end{eqnarray}
Hence the gap at ${\bf k'} = 0$ vanishes at $J_2 = J_1/2$, regardless of
$J_3$, corresponding to a phase transition.

\subsection{Columnar Phase}

There are again two sublattices, with the A and B sites forming
successive columns. The Fourier transformed Hamiltonian is now
\begin{widetext}
\begin{eqnarray}
H & = & -NS^2(2J_2+J_3) +2S\sum_k (J_1 \cos k_y +2J_2  + J_3)(a^{\dagger}_ka_k +
b^{\dagger}_kb_k) \nonumber \\
 & & +2S\sum_k (J_1 \cos k_x +2J_2 \mu_k + J_3 \nu_k)
(a^{\dagger}_kb^{\dagger}_k + a_kb_k) 
\end{eqnarray}
\end{widetext}
A Bogoliubov transformation then yields
\begin{equation}
H = E_0 + \sum_k \omega_k (A^{\dagger}_kA_k +
B^{\dagger}_kB_k) 
\end{equation}
with
\begin{equation}
E_0 = -NS(S+1)(2J_2+J_3) + \sum_k \omega_k
\end{equation}
and
\begin{eqnarray}
\omega_k & = & 2S\sqrt{P^2_k - Q^2_k} \nonumber \\
P_k & = & J_1  \cos k_y + 2J_2 + J_3  \nonumber \\
Q_k & = & J_1 \cos k_x + 2J_2\cos k_x \cos k_y \nonumber \\
 & & + J_3 \cos k_z
\end{eqnarray}
The sublattice magnetization is again given by the formula (\ref{eq14}).

For small $k$
\begin{eqnarray}
P_k - Q_k & = & \frac{1}{2}(J_1 + 2J_2)k^2_x +\frac{1}{2}(2J_2 -
J_1) k^2_y \nonumber \\ 
& & +      
\frac{1}{2}J_3 k^2_z + \cdots 
\end{eqnarray}
which shows that again there is a Goldstone mode at $\bf{k} = 0$, which
is stable for $J_2 > J_1/2$, but
becomes unstable at $J_2 = J_1/2$.

Another special case is ${\bf k} = (k_x,\pi,0)$, when 
\begin{equation}
P_k -Q_k = (2J_2-J_1)(1+\cos k_x) + \cdots .
\end{equation}
Hence $\omega_k$ vanishes along this line at $J_2 = J_1/2$,
corresponding again to the transition point.


\begin{thebibliography}{11}

\bibitem{chandra}P. Chandra and B. Doucot, Phys. Rev. B \textbf{38},
9335 (1988).

\bibitem{dagotto} E. Dagotto and A. Moreo, Phys. Rev. Lett. {\bf 63}, 2148 (1989).

\bibitem{schulz} H.J. Schulz and T.A.L. Ziman, Europhys. Lett. {\bf 18}, 355 (1992); H.J.
Schulz, T.A.L. Ziman and D. Poilblanc, J. Phys. {\bf I6}, 675 (1996).
\bibitem{richter} J. Richter, Phys. Rev. B{\bf 47}, 5794 (1993); J. Richter, N.B. Ivanov and K.
Retzlaff, Europhys. Lett. {\bf 25}, 545 (1994).
\bibitem{oweihong}J. Oitmaa and Zheng Weihong, Phys. Rev. B \textbf{54},
3022 (1996).

\bibitem{bishop} R.F. Bishop, D.J.J. Farnell and J.B. Parkinson, Phys. Rev. B{\bf 58}, 6394
(1998).
\bibitem{singh} R.R.P. Singh, Zheng Weihong, C.J. Hamer and J. Oitmaa, Phys. Rev. B{\bf 60},
7278 (1999).
\bibitem{capriotti}L. Capriotti and S. Sorella, Phys. Rev. Lett.
\textbf{84}, 3173 (2000).

\bibitem{siurak} L. Siurakshina, D. Ihle and R. Hayn, Phys. Rev. B{\bf 64}, 104406 (2001).
\bibitem{sushkov}O. P. Sushkov, J. Oitmaa and Z. Weihong, Phys. Rev.
B \textbf{63}, 104420 (2001).
\bibitem{capriotti3} L. Capriotti, F. Becca, A. Parola and S. Sorella, Phys. Rev. Lett. {\bf
87}, 097201 (2001).

\bibitem{capriotti2}L. Capriotti, F. Becca, A., Parola, A., and S. Sorella, Phys.
Rev. B \textbf{67}, 212402 (2003).
\bibitem{singh2} R.R.P. Singh, Weihong Zheng, J. Oitmaa, O.P. Sushkov and C.J. Hamer, Phys.
Rev. Lett. {\bf 91}, 017201 (2003).
\bibitem{roscilde} T. Roscilde, A. Feiguin, A.L. Chernyshev, S. Liu and S. Haas, Phys. Rev.
Lett. {\bf 93}, 017203 (2004).

\bibitem{sirker} J. Sirker, Z. Weihong, O. P. Sushkov and J. Oitmaa,
Phys. Rev. B \textbf{73}, 184420 (2006).

\bibitem{schmalfuss} D. Schmalfuss, R. Darradi, J. Richter, J. Schulenberg
and D. Ihle, Phys. Rev. Lett. {\bf 97}, 157201 (2006).

\bibitem{ohweihong}J. Oitmaa, C. J. Hamer and Z. Weihong, \textit{\textcolor{black}{Series
Expansion Methods for Strongly Interacting Lattice Models}}, Cambridge,
(2006)

\bibitem{Rosner}H. Rosner, R. R. P. Singh, W. H. Zheng, J. Oitmaa,
S.-L. Drechsler and W. E. Pickett, Phys. Rev. Lett. \textbf{88},
186405 (2002); H. Rosner, R. R. P. Singh, W. H. Zheng, J. Oitmaa,
and W. E. Pickett, Phys. Rev. B \textbf{67}, 014416 (2003).

\bibitem{Uhrig}G. S. Uhrig, M. Holt, J. Oitmaa, O. P. Sushkov, and
R. R. P. Singh, Phys. Rev. B \textbf{79}, 092416 (2009).

\bibitem{Holt}M. Holt, O.P. Sushkov, D. Stanek and G.S. Uhrig,
preprint arXiv:1010.5551v1 (2010)

\bibitem{huse1988} D. A. Huse, Phys. Rev. B {\bf 37}, 2380 (1988).
\bibitem{zheng1991} Zheng Weihong, J. Oitmaa and C.J. Hamer, Phys. Rev.
B \textbf{43}, 8321 (1991).

\bibitem{singh1993} R.R.P. Singh, Phys. Rev. B{\bf47}, 12337 (1993);
R.R.P. Singh and M.P. Gelfand, {\it ibid.} {\bf52}, R15695 (1995).

\bibitem{nunes} W. A. Nunes, J. Ricardo de Sousa, J. Roberto Viana
and J. Richter, J. Phys. Cond. Mat. {\bf 22}, 146004 (2010).

\bibitem{majumdarnew} K. Majumdar, J. Phys. Cond. Mat. {\bf 23}, 046001
(2011).

\bibitem{majumdar}K. Majumdar and T. Datta, J. Stat. Phys. \textbf{139},
714 (2010).
\end{thebibliography}
\end{document}